# Characterizing and Mitigating Self-Admitted Technical Debt in Build Systems

Tao Xiao [ID], Dong Wang [ID], Shane McIntosh [ID], Hideaki Hata [ID], Raula Gaikovina Kula [ID], Takashi Ishio [ID], *Member, IEEE*, and Kenichi Matsumoto [ID], *Senior Member, IEEE*

**Abstract**—Technical Debt is a metaphor used to describe the situation in which long-term software artifact quality is traded for short-term goals in software projects. In recent years, the concept of self-admitted technical debt (SATD) was proposed, which focuses on debt that is intentionally introduced and described by developers. Although prior work has made important observations about admitted technical debt in source code, little is known about SATD in build systems. In this paper, we set out to better understand the characteristics of SATD in build systems. To do so, through a qualitative analysis of 500 SATD comments in the Maven build system of 291 projects, we characterize SATD by location and rationale (reason and purpose). Our results show that limitations in tools and libraries, and complexities of dependency management are the most frequent causes, accounting for 50% and 24% of the comments. We also find that developers often document SATD as issues to be fixed later. As a first step towards the automatic detection of SATD rationale, we train classifiers to detect the two most frequently occurring reasons and the four most frequently occurring purposes of SATD in the content of comments in Maven build systems. The classifier performance is promising, achieving an F1-score of 0.71–0.79. Finally, within 16 identified 'ready-to-be-addressed' SATD instances, the three SATD submitted by pull requests and the five SATD submitted by issue reports were resolved after developers were made aware. Our work presents the first step towards understanding technical debt in build systems and opens up avenues for future work, such as tool support to track and manage SATD backlogs.

✦

## 1 INTRODUCTION

THROUGHOUT the software development process, stakeholders strive to build functional, maintainable, and high-quality software. Despite their best efforts, developers inevitably encounter situations where suboptimal solutions, known as *Technical Debt* (TD) are implemented in a software project [9]. Although studies have traced evidence of TD in source code, TD covers a range of software artifacts and processes (i.e., architecture, build, defects, design, documentation, infrastructure, people, process, requirements, service, and testing) [2]. Clear evidence of TD is at the core of self-admitted technical debt (SATD), where developers record the reasoning behind such suboptimal solutions. Potdar and Shihab [38] observed that SATD existed in 31% of source code files.

Although prior work on SATD in source code has made important advances, modern software development has a

• Tao Xiao, Dong Wang, Raula Gaikovina Kula, Takashi Ishio, and Kenichi Matsumoto are with the Nara Institute of Science and Technology, Ikoma, Nara 630-0192, Japan. E-mail: {tao.xiao.ts2, wang.dong.vt8, raula-k, ishio, matumoto}@is.naist.jp.
• Shane McIntosh is with the Cheriton School of Computer Science, University of Waterloo, Waterloo, ON N2L 3G1, Canada. E-mail: shane.mcintosh@uwaterloo.ca.
• Hideaki Hata is with Shinshu University, Matsumoto, Nagano 390-8621, Japan. E-mail: hata@shinshu-u.ac.jp.

Manuscript received 19 Feb. 2021; revised 21 Sept. 2021; accepted 22 Sept. 2021. Date of publication 0 . 0000; date of current version 0 . 0000.
This work was supported by JSPS KAKENHI under Grants JP18KT0013, JP18H04094, JP20K19774, and JP20H05706.
(Corresponding author: Tao Xiao.)
Recommended for acceptance by R. Kazman.
Digital Object Identifier no. 10.1109/TSE.2021.3115772

broader scope than solely producing source code. Indeed, a complex collection of other software artifacts and tools is needed to produce official software releases. At the heart of these, other artifact is the *build system*, which orchestrates tools (e.g., automated test suites, containerization tools, external and internal dependency management) into a repeatable (and ideally incremental) process. Software teams use *build system specifications* to express dependencies within and among internal and external software artifacts. We hypothesize that build systems may also suffer from TD.

To the best of our knowledge, there have been no previous studies that investigate SATD in build specification files. As suggested by the study on SATD between industry and open-source system developers, researchers should expand studies of SATD beyond the source code [51]. To fill this gap, in this paper, we propose to analyze build files and their existing SATD. More specifically, we set out to characterize SATD, explore its potential for automation and evaluate SATD mitigation strategies. By analyzing the 500 SATD comments extracted from 291 GitHub repositories that utilize the Maven build system, we address the following three research questions:

*(RQ1) What are the Characteristics of SATD in Build Systems?*
*Motivation.* It is unclear what characteristics of SATD in build systems exist. Analyzing SATD characteristics (location and rationale) will lay the foundation for understanding the scope of the SATD problems in build systems. In this research question, we analyze the locations of SATD, the reasons for SATD to occur, and the purposes that SATD serve. Tasks like SATD management often require experienced technical stakeholders. Answering this question will lower the barrier to entry for newcomers who take over a legacy system, and non-technical stakeholders who are responsible for decision making in the context of build system maintenance (e.g., project managers). Furthermore, it will lay the foundation for future research and tool development on SATD problems of practical relevance.

*(RQ1.1) Location: Which Build File Specifications are Most Susceptible to SATD?*

*Results.* SATD tends to occur in the plugin configuration and the external dependencies configuration code, accounting for 49% and 32%, respectively.

*(RQ1.2) Rationale: What Causes a Developer to Document SATD and What Purpose Does it Serve?*

*Results.* We analyze rationale along reason and purpose dimensions. First, we find that there are nine categories of SATD reasons. The most frequent reasons include the limitations in tools and libraries, and complexities of dependency management, accounting for 50% and 24% of analyzed SATD instances, respectively. Second, we find that there are six purposes for leaving SATD comments, with documenting issues to be fixed later and explaining the rationale for a workaround occurring the most frequently, accounting for 39% and 22% of analyzed SATD instances, respectively.

*(RQ2) Can Automated Classifiers Accurately Identify the Characteristics of SATD in Build Systems?*

*Motivation.* The qualitative approach that we used to address RQ1 is not scalable enough for large-scale analyses of SATD categories in build files. Moreover, analysis of build systems at large organizations like Google would require an automated approach to be practical [35]. For practitioners, automatic SATD identification could facilitate the replication of detection approaches to promote their adoption, and improve the detection quality and traceability. Hence, we explore the feasibility of training automatic classifiers to identify SATD characteristics.

*Results.* Experimental results show automation is feasible, achieving a precision of 0.72 and 0.81, a recall of 0.71 and 0.80, and an F1-score of 0.71 and 0.79 for SATD reasons and purposes, respectively. Comparing both traditional and state-of-the-art machine learning techniques, we find that the auto-sklearn based classifiers achieve the highest value over the set of baseline classifiers, i.e., Naive Bayes (NB), Support Vector Machine (SVM), and k-Nearest Neighbor (kNN) by a margin of 16–25 percentage points in terms of average F1-score.

*(RQ3) To What Extent can SATD be Removed in Build Systems?*

*Motivation.* The manner by which developers handle SATD is currently unknown, i.e., whether or not they can be removed. Hence, we investigate the 'ready-to-be-addressed' SATD removal by submitting pull requests and issue reports. Answering this research question can address the necessity of proposing automatic tools to identify SATD comments for researchers and furthermore facilitate better technical debt management for projects.

*Results.* Within 16 'ready-to-be-addressed' SATD instances, we propose pull requests for seven cases, three of which were merged. Moreover, we produce issue reports for nine cases, five of which were resolved within 20 days. While there are a number of factors at play, these responses suggest that developers are receptive and reactive to SATD.

*Replication Package.* To facilitate replication and future work in the area, we have prepared a replication package,

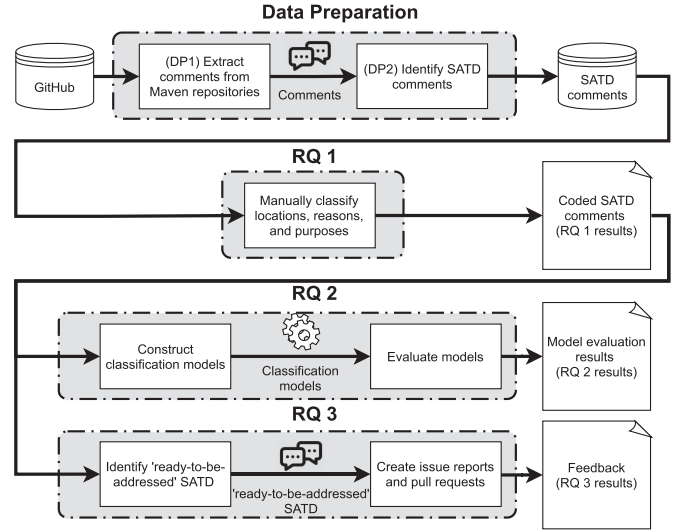

Fig. 1. Overview of the study.

which includes raw data, the manually labelled dataset, and the scripts for reproducing our analyses. The package is available online at https://github.com/NAIST-SE/SATDinBuildSystems.

The remainder of the paper is organized as follows. Section 2 describes the workflow that we followed to collect SATD comments. Sections 3, 4, and 5 show the experiments that we conducted to address RQ1–3, respectively. Section 6 presents the developer survey to obtain developer perceptions and evaluate the SATD reason and purpose categories. Section 7 discusses the recommendations for build system stakeholders based on our study results. Section 8 situates our work with respect to the literature on build systems and technical debt. Section 9 discusses the threats to validity. Section 10 draws the conclusions and highlights opportunities for future work.

## 2 DATA PREPARATION

In this section, we describe the data collection procedure. Fig. 1 shows an overview of the procedure, which consists of two steps: (DP1) Extract comments from Maven repositories; and (DP2) Identify SATD comments.

*(DP1) Extract Comments From Maven Repositories.* Maven is a popular build automation tool used primarily for Java projects. In a large-scale analysis of 177,039 repositories, McIntosh *et al.* [33] found that Maven repositories tend to be most actively maintained. Since developers are actively updating their Maven files, we suspect that technical debt may also be accruing. Thus, we select Maven as our studied build system.

We analyze Java repositories in the dataset shared by Hata *et al.* [16]. That dataset includes the Git repositories of actively developed software projects containing (i) more than 500 commits; and (ii) at least 100 commits during the most active two-year period. Forked repositories are excluded from the analysis. We analyze the latest version (HEAD revision) of the repositories. The list of HEAD revisions is summarized in the replication package.

We select the Maven specifications from each studied repository using the filename convention (i.e., *pom.xml*).

TABLE 1
Summary of Studied Dataset

| # Maven repositories | # POM files | # Comments | # SATD comments |
|---|---|---|---|
| 3,710 | 100,765 | 253,555 | 3,424 |

Each studied repository may have several Maven specifications. Since the specifications are written in XML, comments are recognized as content appearing between "<!--" and "-->" XML tokens. We extract comment content from Maven specifications using a script that builds upon the Java SE XML parser.[1] Finally, we extract 253,555 comments from 100,765 POM files in 3,710 Maven repositories.

*(DP2) Identify SATD Comments*. We detect SATD comments using the keyword-based approach of Potdar and Shihab [38]. In addition, to reduce the risk of missing SATD comments and enlarge the dataset, we expand Potdar and Shihab's keyword list to include 13 frequent features that were recommended by Huang *et al*. [18]. Our adapted list of SATD keywords is summarized in our online appendix.[2] In the end, we are able to detect 3,424 SATD comments. Table 1 provides an overview of our studied dataset.

To evaluate whether or not the set of keywords being used for retrieving SATD comments is accurate enough, we conduct a qualitative study of a statistically representative and stratified sample of comments that are not identified as SATD by the keyword-based approach. The size of the sample comments is calculated to achieve estimates with a confidence level of $95\%$ and a confidence interval of $\pm 5\%$. In total, we have randomly sampled 384 comments from 248,502 comments that are not identified as SATD by the keyword-based approach. Then the first two authors manually checked whether or not they are SATD. In the end, we identified that only thirteen comments are SATD, accounting for 3.4% of comments missing from the keyword-based approach.

## 3 CHARACTERIZING SATD COMMENTS (RQ1)

SATD comments may appear in different locations within build files. Moreover, the rationale for incurring SATD may differ. In this section, we analyze the locations, reasons for adoption, and purposes served by SATD comments. To perform our analyses, we use a manually-intensive method. Below, we present our approach to classify SATD comments according to locations, reasons, and purposes (Section 3.1), followed by the results (Section 3.2). Finally, we explore the relationships between locations and reasons, and locations and purposes (Section 3.3).

### 3.1 Approach

We apply an open coding approach [8] to classify randomly sampled SATD comments in build files. Open coding is a qualitative data analysis method by which artifacts under inspection (SATD comments in our case) are classified according to emergent concepts (i.e., codes). After coding, we apply open card sorting [36] to raise low-level codes to higher levels and more abstract concepts, especially for SATD reasons. This approach involves the first three authors of this paper (the first two authors are graduate students and the third author has more than ten years of research and three years of industrial experience on build systems). Below, we describe our code saturation, sample coding, and card sorting procedures in more detail.

*Code Saturation*. Section 2 shows that there are 3,424 SATD comments out of 253,555 comments appearing in the curated set of GitHub Maven repositories. Since coding of all 3,424 instances is impractical, we elect to code a sample of SATD comments.

First, we initiate a set of codes within the first 50 comments. To make as complete of a list of SATD locations, reasons, and purposes as possible, we strive for *theoretical saturation* [11] to achieve analytical generalization. Similar to prior work [17], we set our saturation criterion to 50, i.e., the first two authors continue to code randomly selected SATD comments until no new codes have been discovered for 50 consecutive comments. Then, they open discussions on classifying SATD comments in terms of locations, reasons, and purposes and try to reach a consensus on disagreements between them. During these discussions, the third author resolves each disagreement and suggests possible improvements on the categories. Finally, we reach saturation after coding 266 SATD comments. Since codes that emerge late in the process may apply to earlier reviews, we performed another pass over all of the comments to correct miscoded entries and tested the level of agreement of our constructed codes. We calculate the Kappa agreement of our codes among the first two authors, who independently coded the locations, reasons, and purposes of all 266 SATD comments. Cohen's Kappa for location codes is 0.91 or 'Almost perfect' agreement [48], whereas Cohen's Kappa for reason and purpose codes are 0.78 and 0.75, respectively, which indicate 'Substantial' agreement [48]. The somewhat lower agreement can be explained by the need for extrapolation when coding the reason and purpose of an instance of SATD from its context and comment content.

*Sample Coding*. To increase the generalizability of our results, after our codes achieve saturation, we coded additional SATD comments to reach 500 samples. We divided the additional 234 samples into two sets (i.e., 117 for each set). Inspired by the encouraging Kappa agreement, the first author independently coded the first set, and the second author independently coded the second set. When coding each SATD comment, the coders focus on the location, key reason, and key purpose. For example, a SATD comment from the `Apache OODT` project is located in the plugin configuration. The reason of this comment is labelled as 'External library limitation' and its purpose is 'Document for later fix'. In a total of 500 samples, the coding process took 120 person-hours.

Since the open coding is an exploratory data analysis technique, it may be prone to errors. To mitigate errors, we code in two passes. First, we code based on the comment itself. After completing an initial round of coding, we perform a second pass over all of the SATD comments to correct miscoded instances. In the second pass, we code based on contextual information, such as the surrounding build specification code, prior commit history, and other relevant development records. Using only the comment content shown in the example below, we could not identify why this SATD occurred. However, using the contextual information surrounding the comment and the information provided in the hyperlink, we could identify the reason. After the second pass, 41 instances are corrected based on the contextual information.

---

1. https://github.com/takashi-ishio/CommentLister
2. https://doi.org/10.6084/m9.figshare.13018580

TABLE 2
Definition and Frequency of SATD Locations and Lines of Code (LOC) of Build Code

| Category | Definition | Frequency | | LOC | |
|---|---|---|---|---|---|
| Plugin configuration | Build code that specifies which build plugin features should be included or excluded and how they should be configured for build execution, e.g., `<plugins>`, `<profiles>`. | 244 | (49%) | 94,377 | (43%) |
| External dependencies configuration | The system under construction or the plugins in use during the build process may rely on external code in order to function correctly. The Maven build technology also includes tooling for specifying and resolving these dependencies through a central repository of artifacts. Maven users may declare their external dependencies in Maven specifications using tags, e.g., `<dependencies>`. | 159 | (32%) | 97,797 | (44%) |
| Build variables | Maven users may declare their own build variables or override inherited variables from a parent POM, e.g., `<properties>`. | 57 | (11%) | 7,689 | (3%) |
| Multi-directory configuration | Build code that avoids redundancies or duplicate configurations through inheritance. Maven users may declare their project's parent in Maven specifications using tags, e.g., `<parent>`. | 10 | (2%) | 5,325 | (2%) |
| Resource configuration | Build code that specifies where resources are stored and what kinds of resources are used during the build process, e.g., `<resources>`. | 10 | (2%) | 1,276 | (1%) |
| Repository configuration | The system which is needed to deploy artifacts from the organization may rely on remote repositories in order to populate the required dependencies to a local repository, e.g., `<repositories>`. | 9 | (2%) | 4,076 | (2%) |
| Project metadata | Build code that specifies project descriptive information. In Maven, this information includes the version, artifact, and group identifiers of the project, e.g., `<artifactId>`, `<groupId>`, `<url>`. | 6 | (1%) | 8,715 | (4%) |
| Build organization | Build code that specifies the build lifecycle and its outputs should be configured, e.g., `<packaging>`. | 3 | (1%) | 414 | (0%) |
| Software configuration management | Build code that specifies a set of information for release build to check out the tag that was created for this release. Maven users may declare it in Maven specifications using tags. e.g., `<scm>`. | 2 | (1%) | 787 | (0%) |

Example: Comment Content

```
<!-- Fix Eclipse Maven integration -->
```

Example: Contextual Information Surrounding the Comment

```
<!-- Since its 1.0 version, the Eclipse Maven integration requires this info to
map several workspace actions to specific phases of Maven's lifecycle.
Should this section not be included, or should certain plugins not be
covered by the list of "pluginExecutions" below, Eclipse users will see
errors of the type "plugin execution not covered". -->
<!-- See https://www.eclipse.org/m2e/documentation/m2e-execution-not-
covered.html for more information -->
```

*Card Sorting.* We apply open card sorting to construct a taxonomy of SATD reasons. Open card sorting helps us to generate general categories from our low-level codes. The open card sorting includes two steps. First, the coded comments are merged into cohesive groups that can be represented by a similar subcategory. Second, the related subcategories are merged to form categories that can be summarized by a short title.

## 3.2 Results

In this section, we present our results for RQ1, consisting of SATD location and rationale (reason and purpose).

*RQ1.1 - SATD Location*

**Observation 1.** *Plugin configuration and External dependencies configuration are the most frequently occurring locations in our sample.* We identified the nine location codes that emerged from our qualitative analysis. Table 2 provides an overview of the categories and their definitions, frequencies, and lines of code (LOC) for SATD locations. From the table, we observe that *Plugin configuration* is the most frequently occurring location for developers to leave SATD comments, with 49% of SATD comments appearing in that location. The second most frequently occurring location is the *External dependencies configuration* location, accounting for 32% of SATD comments appearing in that location. In addition, locations such as *Project metadata*, *Build organization*, and *Software configuration management* are rarely associated with SATD, i.e., 1% for each category.

The fourth column of Table 2 shows that our location tendencies appear to follow the volume of code in each category. This result shows that, as one might expect, categories that require larger volumes of build configuration code tend to be more prone to contain SATD.

*RQ1.2 - SATD Rationale*

*Reasons.* The top portion of Table 3 defines and quantifies the reasons that we observe for SATD comments in our studied sample.

**Observation 2.** *Limitation is the most common reason category for developers to leave SATD comments.* We identified 16 subcategories that emerged from our classification for SATD reasons. The 16 subcategories fit into nine categories. Table 3 shows definitions and frequencies for various SATD reason categories. As we can see from the table, 50% of SATD comments are left due to the *Limitation* reason.

Upon closer inspection, external library limitation is the main limitation, accounting for 41% of the occurrences of the Limitation category. Indeed, it appears that working around the constraints imposed by external libraries is a complexity of modern development from which build specifications are not exempt. The following is an example of the *Limitation* reason. The comment describes the limitation of the specific version of the `maven-war-plugin` plugin.[3]

```
<!-- This is broken in maven-war-plugin 2.0, works in 2.0.1 -->
<warSourceExcludes>WEB-INF/no-lib/*.jar</warSourceExcludes>
```

The second most frequently occurring category is the *Dependency* reason (24%). In an example below, the `bval-jsr` dependency is missing on `jaxb-api`. Thus, developers leave comments as a reminder for why this dependency is needed.[4]

```
<dependency>
    <groupId>org.apache.bval</groupId>
    <artifactId>bval-jsr</artifactId>
</dependency>
<!--Java 9 workaround for missing bval-jsr dependency declaration BVAL-155
-->
<dependency>
    <groupId>javax.xml.bind</groupId>
    <artifactId>jaxb-api</artifactId>
</dependency>
```

Least frequently occurring patterns include the *Code smell* reason (1%) and the *Change propagation* reason (1%). Finally, only 3% of comments are labelled as *No reason*, which means that we could not determine the reasons from them.

In the study of Mensah *et al.* [34], they identified the possible causes of SATD introduction. The most prominent causes are code smells (23%), complicated and complex tasks (22%), inadequate code testing (21%), and unexpected code performance (17%). Comparatively, those causes account for only 1% of the SATD. We suspect that this is because build code specifies a set of rules to prepare and transform the source code into deliverables. Unlike imperative or object-oriented systems, build specifications are primarily implemented to inform an expert system so that it may make efficient and correct decisions. This change in paradigm is likely changing the characteristics of observable SATD.

We describe the remaining reason categories in detail using representative examples in our online appendix[5] to help the reader understand this taxonomy.

*Purposes.* The bottom portion of Table 3 defines and quantifies the purposes that we observe for SATD comments in our studied sample.

**Observation 3.** *Document for later fix is the most frequently occurring purpose.* Our classification revealed six SATD purposes. Table 3 shows the results of our purpose classification. We observe that 39% SATD comments are left by developers with the *Document for later fix* purpose. The result indicates that SATD comments are likely to be used as a short-term memo for developers to recheck in the future. For example, since Maven resolves dependencies transitively, it is possible to include unwanted or problematic dependencies. For the `software.amazon.awssdk:s3` dependency which

TABLE 3
Definition and Frequency of SATD Reasons and Purposes

| | Category | Definition | Frequency | |
|---|---|---|---|---|
| Reason | **Limitation** | Constraints imposed by the design or implementation of third-party libraries or development tools. | **249** | **(50%)** |
| | External library limitation | | 206 | (41%) |
| | External tool limitation | | 27 | (5%) |
| | Build tool limitation | | 16 | (3%) |
| | **Dependency** | Dependency issues due to unavailable artifacts or assets, such as missing or stale dependencies, dependency conflicts, or dependency resolution in post-install phase. | **123** | **(24%)** |
| | Stale dependency | | 67 | (13%) |
| | Missing dependency | | 47 | (9%) |
| | Dependency conflict | | 5 | (1%) |
| | Post-install dependency resolution | | 4 | (1%) |
| | **Recursive call** | Coherence issues, recursive calls to invoke another build file. | **40** | **(8%)** |
| | **Document** | Inadequate project description issues, such as licensing and metadata specification. | **23** | **(5%)** |
| | Specify metadata | | 12 | (2%) |
| | Licensing | | 11 | (2%) |
| | **Build break** | Broken builds (i.e., failures that occur during the build process) in build files. | **20** | **(4%)** |
| | **Compiler setting** | Configuration issues during the compilation process, such as compiler configuration and symbol visibility. | **19** | **(4%)** |
| | Compiler configuration | | 17 | (3%) |
| | Symbol visibility | | 2 | (1%) |
| | **Code smell** | Violations of fundamentals of design principles, i.e., instances of poor coding practice in build files. | **7** | **(1%)** |
| | **Change propagation** | Changes that need to be propagated to keep software artifacts in sync during updates. | **2** | **(1%)** |
| | **No reason** | A label could not be assigned (due to lack of information). | **17** | **(3%)** |
| Purpose | Document for later fix | Document an issue that should be revisited in the future. | 194 | (39%) |
| | Document workaround | Explicitly document constraints imposed by design or implementation choices. The comment contains workaround-related keywords, such as "workaround" and "temporary." | 110 | (22%) |
| | Warning for future developers | Warn other developers to pay attention to an aspect of the solution that may not be clear from its structure or content. | 95 | (19%) |
| | Document suboptimal implementation choice | Explain why a problematic solution has been adopted. | 79 | (16%) |
| | Placeholder for later extension | Document an extension point for later enhancement(s). | 13 | (3%) |
| | Silence build warnings | Defer or ignore warnings emitted by underlying tools. | 9 | (2%) |

*Nine categories merged from subcategories for SATD reasons are shown in bold.*

includes the broken `netty-nio-client:software.amazon.awssdk` dependency, developers exclude this dependency to preserve a clean (i.e., non-broken) build status. Developers leave this comment as a note to revisit in the future.[6]

```
<dependency>
    <groupId>software.amazon.awssdk</groupId>
    <artifactId>s3</artifactId>
    <version>${awsjavasdk.version}</version>
    <exclusions>
        <exclusion>
            <!-- TODO remove exclusions after we fix netty module -->
            <artifactId>netty-nio-client</artifactId>
            <groupId>software.amazon.awssdk</groupId>
        </exclusion>
    </exclusions>
</dependency>
```

Another commonly occurring purpose is the *Document workaround* purpose, accounting for 22% of our sample. In the example below, the `io.grpc:grpc-core` dependency is partly unusable. Developers comment out this dependency and leave the comment to document this temporary fix.[7]

---

3. https://tinyurl.com/y5jtkxxb
4. https://tinyurl.com/ymvvfdss
5. https://doi.org/10.6084/m9.figshare.13147727

6. https://tinyurl.com/y4wg8n3z
7. https://tinyurl.com/y43rxj9a

```
<!-- FIXME(lesv) Temporary fix due to Datastore having the wrong version
    -->
    <!--<dependencyManagement>
        <dependencies>
            <dependency>
                <groupId>io.grpc</groupId>
                <artifactId>grpc-core</artifactId>
                <version>1.2.0</version>
            </dependency>
        </dependencies>
    </dependencyManagement>-->
<!-- end of FIXME -->
```

On the other hand, only 3% and 2% of SATD comments are identified for the *Placeholder for later extension* purpose and the *Silence build warnings* purpose.

The survey of Maldonado *et al.* [28] showed that developers most often use SATD to track future bugs and bad implementation areas. In our context of build systems, the high frequency of the *Document for later fix* purpose agrees with their observations.

We describe the remaining purpose categories in detail using representative examples in our online appendix[8] to help the reader understand this taxonomy.

### 3.3 Relationships

**Observation 4.** *Location categories share a strong relationship with reason categories.* Motivated by representative examples, we observe that SATD comments in similar locations can vary in terms of reasons and purposes. Thus, we conduct a further study to investigate the relationships between locations and reasons, and locations and purposes. We visualize these relationships by using two parallel sets [21] in parallel categories diagrams. Parallel sets are variants of parallel coordinates, in which the width of lines that connect sets corresponds to the frequency of their co-occurrence. Fig. 2 shows relationships between locations and reasons, and locations and purposes.

In Fig. 2a, SATD comments in *Plugin configuration* most frequently occur (66.0%) because of the *Limitation* reason in our sample. The example below shows a co-occurrence of *Plugin configuration-Limitation*.[9] This SATD comment is located in the `<plugin>` tag and indicates that the current plugin suffers from an external library limitation, i.e., `maven-shade-plugin`.

```
<plugin>
    <artifactId>maven-antrun-plugin</artifactId>
    <executions>
        <!-- XXX: workaround for an issue with maven-shade-plugin
        There appears to be some stale state from previous executions of the
            Shade plugin, which manifest themselves as "We have a duplicate
            " warnings on build and also as some classes not being updated
            on build. -->
        <execution>
...
</plugin>
```

SATD comments located in *External dependencies configuration* most frequently occur (60.4%) due to the *Dependency* reason in our sample. The example below shows this relationship, where a comment located in the

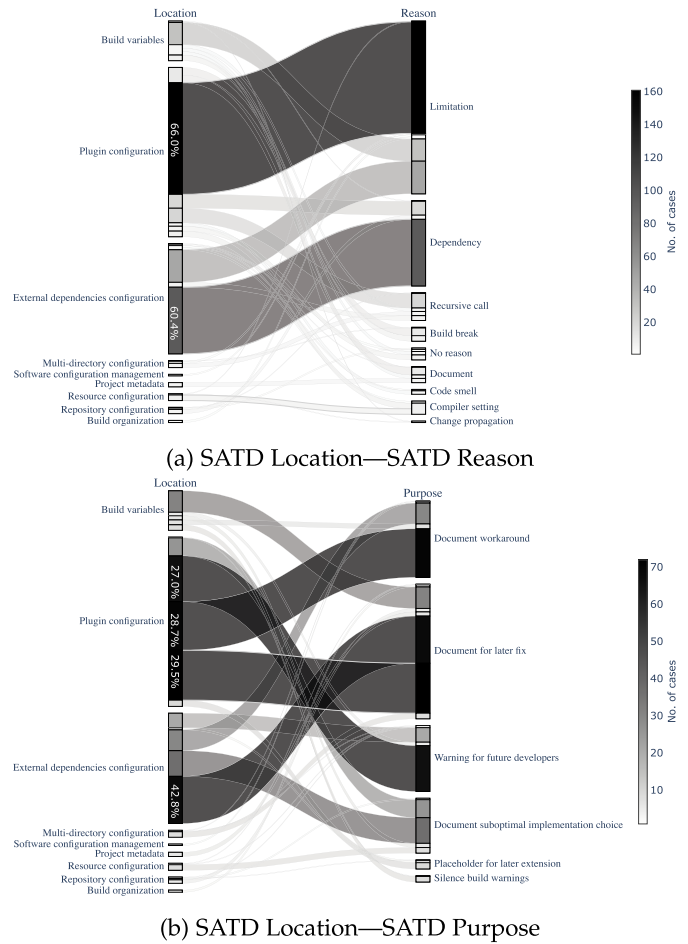

(a) SATD Location—SATD Reason

(b) SATD Location—SATD Purpose

Fig. 2. Parallel sets between locations and reasons, and locations and purposes of SATD in Maven build systems. For example, *Plugin configuration* most frequently occur because of the *Limitation* reason.

`<dependencies>` tag indicates that it should adopt the proper dependency (i.e., `org.apache.calcite.avatica:avatica`) due to the inconsistency `protobuf` dependency with Hadoop.[10]

```
<dependencies>
    <!-- It should be avatica(the shaded one), not avatica-core, since the
        inconsistency protobuf dependency with Hadoop -->
    <dependency>
        <groupId>org.apache.calcite.avatica</groupId>
        <artifactId>avatica</artifactId>
        <exclusions>
            <exclusion>
                <groupId>org.apache.calcite.avatica</groupId>
                <artifactId>avatica-core</artifactId>
            </exclusion>
        </exclusions>
    </dependency>
</dependencies>
```

These two observations suggest that location categories tend to be more prone to SATD causes.

Moreover, for the relationships between locations and purposes that are shown in Fig. 2b, SATD comments in the *Plugin configuration* location are most often left with the *Document for later fix* purpose (29.5%). For instance, the SATD comment below describes an issue to be revisited later.[11]

```
<plugin>
    <groupId>org.apache.maven.plugins</groupId>
    <artifactId>maven-antrun-plugin</artifactId>
    <executions>
        <execution>
            <id>download</id>
            <phase>generate-resources</phase>
            <configuration>
                <!--
                TODO: I would like to add an "unless" constraint to the target
                that prevents execution if Maven operates in offline mode
                . However I was not able to find out how to obtain this
                information. ${settings.offline} (as noted by several
                resources) does not work. Until fixed builds will fail if no
                internetconnection is available!
                -->
                ...
</plugin>
```

Fig. 2b also shows that SATD comments in the *External dependencies configuration* location are most often left with the *Document for later fix* purpose (42.8%). In a case below, the comment located in the `<dependency>` tag is used to document a pending `maven-plugin 3.0`.[12] Despite these two observations, there is no clear relationship between locations and purposes of SATD comments. The result suggests that the location of SATD does not affect the purpose of developers leaving SATD comments or vice versa.

```
<dependency>
    <groupId>org.jenkins-ci.main</groupId>
    <artifactId>maven-plugin</artifactId>
    <version>2.6</version>
    <exclusions>
        <exclusion> <!-- TODO pending maven-plugin 3.0 -->
            <groupId>org.apache.ant</groupId>
            <artifactId>ant</artifactId>
        </exclusion>
    </exclusions>
</dependency>
```

> *RQ1 Summary*. We identified nine SATD locations, nine reasons, and six purposes in Maven build systems. Location categories tend to be more prone to SATD causes. In the build system maintenance, stakeholders involved in SATD management should be aware of diverse SATD characteristics to assist in making effective management decisions. For example, using these categories, stakeholders can measure and classify these SATD to calculate priority or estimate damage in the future.

## 4  SATD COMMENT CLASSIFICATION (RQ2)

In Section 3, our results provide evidence that diverse SATD locations and rationale (i.e., reasons and purposes) exist in build files. To facilitate the replication of detection approaches, and promote the adoption and traceability of SATD, an automated SATD classifier should be beneficial.

Thus, we further study the feasibility of automatically classifying SATD comments. To do so, we use the manually coded SATD comments from Section 3 as a dataset. With this dataset, we train classifiers based on machine learning techniques and evaluate their performance. Below, we present our approaches to automated classification (Section 4.1) and model evaluation (Section 4.2), as well as the results to RQ2 (Section 4.3).

### 4.1  Approach

We merge infrequent codes (i.e., those whose frequencies are less than 10% of the sample comments), which do not provide sufficient signal to train a reliable classifier. For the reason categories, we merge *Recursive call*, *Document*, *Build break*, *Compiler setting*, *Code smell*, *Change propagation*, and *No reason* into a single "Other" category. For the purpose categories, we merge *Placeholder for later extension* and *Silence build warnings* into a single "Other" category.

*Text Preprocessing*. An analysis of coded SATD comments revealed that bug report links usually appear in comments. Thus, for all SATD comments, we replace hyperlinks with `abstracturl` by using the regular expression similar to the previous study [27]:

```
https?:\/\/(www\.)?[-a-zA-Z0-9@:%._\+~#=]{2,256}\.[a
    -z]{2,6}\b([-a-zA-Z0-9@:%_\+.~#?&//=]*)}
```

Moreover, to reduce the impact of noisy text in comments, we remove special characters by using the regular expression `[^A-Za-z0-9]+`. Additionally, we apply Spacy[13] to lemmatize words, which accounts for term conjugation. Although it is common practice, we opt to exclude stop words removal, since stop words like 'for' and 'until' convey critical semantics in the context of SATD comments [27].

*Feature Extraction*. We apply the N-gram Inverse Document Frequency (IDF) approach to extract features from the preprocessed text using the N-gram weighting scheme tool [43] with its default setting. N-gram IDF [44] is a theoretical extension of the IDF approach for handling words and phrases of any length. The approach generates a list of all valid N-gram terms, and the strength of their association with the targeted classes excluding *Other*. We remove any term that appears only once in each class. In total, 961 and 1,160 N-gram terms are retrieved for SATD reasons and purposes, respectively.

*Classifier Preparation*. Previous studies [26], [27], [49] reported that classifiers trained by combining N-gram IDF and auto-sklearn machine learning tend to outperform classifiers that are trained with single word features. Heeding their advice, we train our classifiers using auto-sklearn [12], which automatically determines effective machine learning pipelines for classification. Auto-sklearn searches a configuration space of 15 classification algorithms, 14 feature preprocesses, and four data preprocesses for optimal hyperparameter settings. We configure the approach to optimize for the weighted F1-score, with a budget of one hour for each round, and a memory capacity of 32 GB.

### 4.2  Evaluation Setting

To evaluate our classifier, we use common performance measures. The precision is the fraction of SATD comments that are correctly classified. The recall is the fraction of truly

---

11. https://tinyurl.com/4ye7jke8
12. https://tinyurl.com/y49s8kg3

13. https://spacy.io/

TABLE 4
Performance of Classifiers for SATD Reason

|  | Category | auto-sklearn | NB | SVM | kNN |
|---|---|---|---|---|---|
| **Precision** | Limitation | **0.75** | 0.70 | 0.67 | 0.72 |
|  | Dependency | 0.72 | 0.65 | **0.84** | 0.68 |
|  | Other | 0.66 | 0.58 | **0.72** | 0.41 |
|  | Avg. | **0.72** | 0.66 | **0.72** | 0.63 |
| **Recall** | Limitation | 0.78 | 0.76 | **0.89** | 0.59 |
|  | Dependency | **0.65** | 0.53 | 0.41 | 0.29 |
|  | Other | 0.65 | 0.59 | 0.51 | **0.74** |
|  | Avg. | **0.71** | 0.66 | 0.67 | 0.55 |
| **F1-score** | Limitation | **0.76** | 0.72 | **0.76** | 0.64 |
|  | Dependency | **0.67** | 0.57 | 0.53 | 0.39 |
|  | Other | **0.65** | 0.57 | 0.57 | 0.52 |
|  | Avg. | **0.71** | 0.65 | 0.65 | 0.55 |

TABLE 5
Performance of Classifiers for SATD Purpose

|  | Category | auto-sklearn | NB | SVM | kNN |
|---|---|---|---|---|---|
| **Precision** | Document for later fix | **0.75** | 0.66 | 0.61 | 0.64 |
|  | Document workaround | 0.95 | 0.57 | **0.99** | 0.90 |
|  | Warning for future developers | 0.77 | 0.48 | **0.85** | 0.68 |
|  | Document suboptimal implementation choice | **0.78** | 0.41 | 0.74 | 0.66 |
|  | Other | **0.93** | 0.85 | 0.70 | 0.50 |
|  | Avg. | **0.81** | 0.58 | 0.76 | 0.70 |
| **Recall** | Document for later fix | 0.92 | 0.52 | **0.96** | 0.88 |
|  | Document workaround | **0.90** | 0.70 | 0.82 | 0.83 |
|  | Warning for future developers | **0.69** | 0.44 | 0.54 | 0.45 |
|  | Document suboptimal implementation choice | 0.48 | 0.42 | 0.32 | 0.25 |
|  | Other | 0.78 | **0.82** | 0.38 | 0.78 |
|  | Avg. | **0.80** | 0.54 | 0.72 | 0.68 |
| **F1-score** | Document for later fix | **0.82** | 0.57 | 0.75 | 0.74 |
|  | Document workaround | **0.93** | 0.62 | 0.89 | 0.86 |
|  | Warning for future developers | **0.71** | 0.43 | 0.64 | 0.54 |
|  | Document suboptimal implementation choice | **0.58** | 0.40 | 0.43 | 0.35 |
|  | Other | **0.83** | 0.80 | 0.49 | 0.57 |
|  | Avg. | **0.79** | 0.54 | 0.70 | 0.66 |

SATD comments that are classified as such. The F1-score is the harmonic mean of precision and recall.

*Comparison.* To investigate the impact that the choice of classification technique has, we apply Naive Bayes (NB), Support Vector Machine (SVM), and k-Nearest Neighbors (kNN) classification techniques. These classifiers have been broadly adopted in prior studies [18], [50]. Similar to prior work [27], we apply TF-IDF [42] to extract the features for our baseline classifiers.

*Ten-Fold Cross-Validation.* To estimate classifier performance on unseen data, we apply ten-fold cross-validation, which divides the data into ten sets and each set is used for testing while others are used for training. Similar to prior work [27], we use the Stratified ShuffleSplit cross validator of scikit-learn [37]. Since some comments can appear multiple times in different sets, we report the mean of the performance measures over ten rounds.

### 4.3 Results

**Observation 5.** *The auto-sklearn classifier achieves the highest value of F1-score for both reasons and purposes.* Table 4 shows the classifier performance with respect to the reasons for SATD. The table shows that the average precision is 0.72, which is greater than the precision of the NB and kNN classifiers (0.66 and 0.63, respectively). The average recall and F1-score of the auto-sklearn are also greater than baseline classifiers by at least four and six percentage points, respectively. Upon closer inspection, we find that the classification of *Limitation* achieves the best performance when compared with the other two reason categories. For instance, the recall and F1-score for *Limitation* are 0.78 and 0.76, respectively, which are greater than the next best performance, *Dependency*, by a margin of at least nine percentage points.

Table 5 shows the classifier performance with respect to the purposes of SATD. As we can see from Table 5, the auto-sklearn classifier outperforms the baseline classifiers. The average precision, recall, and F1-score are greater than other baseline classifiers by at least five points. Closer inspection of the purpose categories reveals that classifying the *Document workaround* purpose has the best performance, with the F1-score reaching 0.93. Such high performance is possible as there usually exist keywords that explicitly map this category, e.g., 'workaround' and 'temporary'. Moreover, for the other purpose categories, we find that the performance is still promising, e.g., achieving F1-scores of 0.82 and 0.71 for *Document for later fix* and *Warning for future developers* purposes.

On the other hand, we observe that SVM outperforms auto-sklearn in terms of precision. For example, SVM reaches greater precision value of most reasons, i.e., *Dependency* with 0.84 and *Other* with 0.72. Moreover, SVM outperforms auto-sklearn in special categories, i.e., *Document workaround* with precision of 0.99, *Warning for future developers* with precision of 0.85, and *Document for later fix* with recall of 0.96.

In Table 6, we list the most frequently occurring N-gram features. For example, 'workaround for' appeared 55 times for SATD comments with the *Document workaround* purpose, and 'be break' appeared 20 times for SATD comments about *Limitation* reason. To better understand our classifiers, we set out to explore why the auto-sklearn classifier fails to correctly classify 129 SATD comments. To do so, we manually review these SATD comments in search of possible causes for their misclassification. Since we focus on the main reasons and purposes (i.e., accounting for relatively high frequencies), we omit the *Other* categories from this analysis. There is a total of 95 reasons and 53 purposes that are misclassified in ten-fold rounds (19 SATD comments are misclassified in both reason and purpose).

In terms of SATD reason categories, we observe that the classifier tends to mainly misclassify the *Dependency* reason as *Limitation* reason (29 cases). We suspect that this is because contextual information is often needed to differentiate between these two categories. Since our classifier only focuses on the comment content itself, the contextual information is missing. For example, the hyperlink was replaced with "abstrcturl" in the text preprocessing step, which makes the URL and its contents opaque to the classifier. In terms of SATD purpose categories, we observe that the classifier tends to misclassify *Document suboptimal implementation choice* (29 cases) and *Warning for future developers* (23 cases) as *Document for later fix*. We observe that these misclassified cases often contain specific keywords, such as TODO. While TODO is often used to indicate areas for future improvements, it also appears within comments that document suboptimal implementation choices and warn future developers of confusing program behaviour.

TABLE 6
Frequently Occurring N-Gram Features in Each Category

| | Category | N-gram features | Frequency |
|---|---|---|---|
| Reason | **Limitation** | be break | 20 |
| | | basedir | 15 |
| | | classifier | 12 |
| | | clover | 12 |
| | | logback | 12 |
| | **Dependency** | should not | 9 |
| | | servicemix | 7 |
| | | require by | 6 |
| | | be here | 6 |
| | | should not be | 5 |
| Purpose | **Document for later fix** | to the | 19 |
| | | carbon | 15 |
| | | offline | 12 |
| | | like | 9 |
| | | group | 8 |
| | **Document workaround** | workaround for | 55 |
| | | workaround for abstracturl | 19 |
| | | workaround to | 14 |
| | | for java | 7 |
| | | from be | 7 |
| | **Warning for future developers** | classifier | 12 |
| | | clover | 12 |
| | | break in | 10 |
| | | hack alert | 6 |
| | | licensefile | 6 |
| | **Document suboptimal implementation choice** | should be able | 9 |
| | | classpath | 8 |
| | | be here | 6 |
| | | todo why | 6 |
| | | the classpath | 5 |

> *RQ2 Summary.* The auto-sklearn classifier achieves the highest value of F1-score for SATD reasons (0.72 precision, 0.71 recall, 0.71 F1-score) and purposes (0.81 precision, 0.80 recall, 0.79 F1-score).

## 5 SATD REMOVAL (RQ3)

In this section, we investigate the willingness of developers to remove the 'ready-to-be-addressed' SATD that contains resolved bug reports similar to the previous study [26]. To do so, we mine for links in the comments of the manually coded data from Section 3. We then systematically assess whether the SATD is ready to be addressed. This concept is known as 'on-hold' SATD, which is a condition to indicate that a developer is waiting for a certain event to occur elsewhere (e.g., an update to the behaviour of a third-party library or tool), according to the study of Maipradit *et al.* [27]. Below, we first describe our studies of the incidences of 'ready-to-be-addressed' SATD (Section 5.1), and our proposed clean-up pull requests and tracking issue reports (Section 5.2).

TABLE 7
Frequency of Link Target Types in
90 SATD Comments

| Category | Frequency | |
|---|---|---|
| bug report | 81 | (78%) |
| 404 | 10 | (10%) |
| tutorial or article | 6 | (6%) |
| Stack Overflow | 2 | (2%) |
| pull request | 2 | (2%) |
| software homepage | 1 | (1%) |
| forum thread | 1 | (1%) |
| blog post | 1 | (1%) |
| **sum** | 104 | (100%) |

### 5.1 Incidences of Ready-to-be-Addressed SATD

*Identify Ready-to-be-Addressed SATD.* We systematically identify 'ready-to-be-addressed' SATD using the following list of conditions:

Step 1. Extract hyperlinks or issue IDs from the comments. Using regular expressions, we extract 104 links from 90 SATD comments. Then we manually code them based on the link target coding guide of Hata *et al.* [16]. Table 7 shows the link target distribution. We observe that *bug report* is the most frequently occurring (78%) link target in SATD comments.

Step 2. Check link target in SATD comments. We check if the link target is a *bug report* with the status of 'resolved', 'closed', 'verified', or 'completed', and resolution type is set to 'fixed' similar to the previous study [26]. Furthermore, to facilitate the creation of our pull requests and issue reports, we exclude four candidates where: (I) the repository referenced in the SATD comment has been archived; (II) the repository referenced by the issue report in the SATD comment has been archived; (III) the repository referenced in the SATD comment is a mirror repository; (IV) the issue report link in the SATD comment is a 'cross-reference' (e.g., the issue report is referenced to aid in documenting the rationale behind an implementation choice).

**Observation 6.** *Of the 90 SATD comments that contain hyperlinks, 16 contain 'ready-to-be-addressed' SATD. Among the 16, Plugin configuration is the most frequently occurring location, Limitation is the most frequently occurring reason, and Document workaround is the most frequently occurring purpose, i.e., 13, 10, and 12 cases, respectively.* Table 8 shows that we initially identified 27 'ready-to-be-addressed' SATD in our dataset. However, we observed that 10 of the 27 SATD had already been removed by developers. Five SATD were removed because the entire file was deleted. The other five SATD were addressed directly by developers. Additionally, one SATD had been submitted as an issue report by a developer, but it has not been closed.

TABLE 8
Distribution of 'Ready-to-be-Addressed' SATD

| Status | Frequency | |
|---|---|---|
| Existing | 16 | (59%) |
| File deleted | 5 | (19%) |
| Fixed by developers | 5 | (19%) |
| Developers try to fix | 1 | (3%) |

In a research project, we analyzed the build file of repositories looking for comments with ready-to-be-addressed SATDs (self-admitted technical debt) that could be addressed. If the underlying issue is already resolved, the related code or comment can be removed or fixed. As we found an instance of this kind of asynchrony in language/thingml/pom.xml, we decided to report the issue to contribute community.

A comment in language/thingml/pom.xml file claims that it is workaround for eclipse/xtext#1231, and this issue has been closed long ago, they also provided new solution: eclipse/xtext#1231 (comment).

Comment location:

> **ThingML/language/thingml/pom.xml**
> Line 88 in c2d2df9
>
> 88     <!-- workaround https://github.com/eclipse/xtext/issues/

(a) Example of an issue report

According to the link https://issues.apache.org/jira/browse/MJARSIGNER-53, I observed that such issue had already been fixed. And based on the comment "Remove once it's in upstream". This override attribute is not needed.

The override attribute and comment:

> **remoting/pom.xml**
> Lines 74 to 75 in 0860f22
>
> 74     <!-- TODO: Required to override by a version with certi
> 75     <maven-jarsigner-plugin.version>1.4</maven-jarsigner-plu

(b) Example of a pull request

Fig. 3. Examples of created issue report and pull request.

## 5.2 Creation of Pull Requests and Issue Reports

To evaluate the importance of 'ready-to-be-addressed' SATD, we created issue reports and pull requests to the studied projects. When preparing issue reports, we also provide possible solutions for developers to deal with 'ready-to-be-addressed' SATD. Since some SATD comments cannot be resolved by directly adding or removing dependencies or removing comments themselves, we submitted issue reports to inform developers of the existence of such examples of SATD in their systems. Examples of issue reports and pull requests are shown in Figs. 3a and 3b.

**Observation 7.** *For 'ready-to-be-addressed' SATD, removal rates have been reached 43% and 56% in pull requests and issue reports, respectively.* In total, we prepared seven issue reports for nine instances of SATD, since three SATD belong to the same repository. In addition, we prepared seven pull requests for the other seven instances of SATD comments. We found developers actively resolve these pull requests and issue reports within 20 days time frame.

Three of the four responded pull requests have been accepted and merged into the main branch. For instance, one developer responded: *"I merged it, and found and fixed two others of the same type, which would have remained if you had not brought it to our attention. Thanks!"* Only one pull request was rejected because the developer had to consider the plugin version, i.e., *"Thanks for the reminder! Upgrading the plugin in my TODO list for the summer, so I'll look into it shortly. The plugin version must be updated before removing the config."*

For the prepared issue reports, five 'ready-to-be-addressed' SATD were resolved. For instance, one developer left the appreciation: *"Thanks for making us aware of this fact."* On the other hand, two issue reports were rejected: one case is where the developer did not agree that the issue was an instance of technical debt (as shown in our online appendix[14]) and another case is where, due to the system supporting multiple versions, the SATD could not be removed.

> *RQ3 Summary.* We identified 16 instances of SATD that are 'ready-to-be-addressed'. Through our experiment, we propose pull requests for seven cases, three of which were merged. Moreover, we produce issue reports for nine cases, five of which were resolved within 20 days. These responses suggest that developers are receptive and reactive to SATD in build systems.

## 6 DEVELOPER FEEDBACK

To evaluate the discovered SATD reason and purpose categories, we conduct a developer survey on the perceptions of SATD in build files. The goal of our survey is to (i) incorporate developer perceptions on instances of TD, (ii) understand real-world experiences with similar instances of TD, and (iii) evaluate the suitability of the proposed reason and purpose categories. The survey consists of ten SATD cases, which we ask our respondents to label. Respondents are also given a table that provides an overview of our discovered labels and their definitions. The cases in the survey are randomly sampled from our studied data set in a stratified manner, i.e., the frequency of category occurrence in the data set guides the likelihood of cases of that category being selected for our surveys. Respondents are allowed to skip any cases or not answer certain questions.

There are 4,746 developers who have made at least one contribution to the 291 studied Maven repositories on GitHub. To ensure that our participants have sufficient expertise, we filter out candidates with fewer than ten contributions to our studied Maven repositories. We invited the remaining 1,670 developers to participate in our survey. The survey was open from July 9 to August 1, 2021. We received responses from 20 developers. Table 9 presents the overview of the demographics of our survey participants. Most of the respondents have more than ten years of Maven editing experience, and their programming experiences vary from 5 years to 30 years and above. Below, we present our survey results.

*Developer Perspectives on TD Types.* Table 10 shows the outcome of the TD labeling task from our respondents. The table shows that in each of the ten prepared cases, more than 50% of developers agree with the instance of a TD. For instance, in Case 6 below,

```
<!-- Temporary workaround. This should be removed once the Karaf is
     properly configured with the remote repos, so that it can download the
     kie-ci-osgi itself (instead of relying on the outer Maven test build). -->
<dependency>
    <groupId>org.kie</groupId>
    <artifactId>kie-ci-osgi</artifactId>
    <scope>test</scope>
</dependency>
```

14. https://doi.org/10.6084/m9.figshare.15059925

TABLE 9
Demographics of Survey Participants

| Gender | male | 18 |
|---|---|---|
| | prefer not to say | 2 |
| **Programming years** | 5–9 | 3 |
| | 10–14 | 2 |
| | 15–19 | 2 |
| | 20–24 | 6 |
| | 25–29 | 2 |
| | 30 and above | 5 |
| **Maven editing years** | less than 3 | 1 |
| | 3–5 | 1 |
| | 6–9 | 7 |
| | 10 and above | 10 |
| | prefer not to say | 1 |

93% of respondents agree that such a case is an instance of TD. In the survey questions, we also asked the developers whether they have experience with similar examples of TD. The survey results show that for each of the ten cases, at least 58% of developers have had a personal experience with a similar example of TD. Moreover, one developer elaborated on how best to deal with cases of inadequate licensing information, stating that *"Ideally I suppose Maven should do that automatically by picking it up from other files."* Indeed, the potential for automating the management and repair of TD in build systems is great.

*Evaluation of Reason and Purpose Categories.* Table 10 also shows the results of our survey-based evaluation of TD reason and purpose categories. For the reason categories, six cases that were labelled by developers are consistent with our labels. Among the remaining four cases, although there are three cases that were labelled with different categories, the second most frequent reason that developers selected matched with our results. Developers only disagree with Case 3 (shown below), labeling it as *Code smell* instead of *Recursive call.*

```xml
<!-- Parent POM defines common plugins and properties. TODO: use the
      parent when this sample passes checkstyles. See: https://github.com/
      GoogleCloudPlatform/cloud-bigtable-examples/issues/59
<parent>
    <groupId>com.google.cloud</groupId>
    <artifactId>bigtable-samples</artifactId>
    <version>1.0.0</version>
    <relativePath>..</relativePath>
</parent> -->
```

In addition, a respondent explained that some debts are inherited because of decisions made by upstream dependencies: *"We faced similar issues with eclipse link. Sometimes you just document that it sucks and move on. There's not a ton to be done when the upstream library is complex, heavily integrated into your project, still active, and broken."*

For the purpose categories, the survey results show that most labels from developers are consistent with our labels from RQ1. As shown in the table, five cases that were labelled by developers (i.e., the most frequent purposes) exactly matched with our labels. Although the most frequent purposes are not aligned with our labels in the rest of the five cases, the second or third most frequently occurring labels are still consistent.

## 7 RECOMMENDATIONS

Based on our findings, we make the following recommendations for practitioners, researchers, and tool builders. First, we recommend that practitioners:

- *Track SATD by using issue trackers,* as we find that developers have tried to add issue report hyperlinks or issue IDs in tandem with comments. Using only comments to track or manage SATD is still problematic. Indeed, only 90 out of the 500 SATD comments contain hyperlinks. Explicitly referencing related content will improve traceability.
- *Check SATD containing resolved bug reports,* as we identified 16 instances of SATD comments that are ready-to-be-addressed from RQ3. These stale SATD comments could create confusion for anyone inspecting the code.

Even more insights for practitioners would be discovered by the following future research:

- *Further studies of workarounds for SATD.* During the coding process, we observe that one SATD work-around can be used across several repositories. This suggests that the retrieval and curation of workarounds may have broad implications beyond the scope of a specific project, and are thus important and of value for practitioners.
- *Establishing an understanding of SATD in other build systems.* As seen in Tables 2 and 3, we identified nine locations, nine reasons, and six purposes, which could improve the overall understanding of SATD in build systems. Applying the coding guides from RQ1 to other build systems (e.g., make, CMake, or Ant) could help to establish a broader theory of SATD in build systems in general.
- *Improving the classification of SATD in build systems.* In RQ2, we propose automatic classifiers to identify SATD characteristics. Tables 4 and 5 show that our classifiers are promising. The results demonstrate the feasibility of automatic classifiers and lay the foundation for developing a SATD classification system. However, it still has room for performance improvement. We suggest that in future research, researchers evaluate other approaches to improve the SATD classifiers. Based on our qualitative analysis of exploring why our auto-sklearn classifier fails to classify the SATD comments, we suggest that contextual information (e.g., surrounding comment content and issue description extracted from the issue report hyperlink) could be included as an input to support the classification of reason categories. Moreover, we suggest that researchers pay attention to the effect of the "TODO" annotation for automatically classifying SATD purpose categories, since it is implicated in several false positive results of the classifier.

The following directions for future work may yield value for tool builders:

- *Tool support for managing SATD in build systems.* Although we recommend that practitioners use issue report hyperlinks or IDs to track SATD, it could be

TABLE 10
Results of Developer Feedback

| Case | Being TD | Similar experience | Reason Ours | Survey result | Purpose Ours | Survey result |
|---|---|---|---|---|---|---|
| 1 | 10/20 (50%) | 15/20 (75%) | Lim. | **Lim. (8)**, Dep. (4), None of them (4), B.b (2) | War. | **War. (9)**, D.w (5), D.lx (4) |
| 2 | 10/16 (63%) | 15/16 (94%) | Dep. | Lim. (7), **Dep. (5)**, C.s (2) | D.w | **D.w (7)**, D.sic (5), D.lx (2), War. (2) |
| 3 | 14/16 (88%) | 12/17 (71%) | R.c | C.s (7), B.b (3), Lim. (3), None of them (2) | D.lx | **D.lx (10)**, D.w (3), D.sic (2) |
| 4 | 11/20 (55%) | 14/19 (74%) | Doc. | **Doc. (13)**, None of them (4) | Pla. | D.lx (10), **Pla. (7)** |
| 5 | 12/17 (71%) | 13/16 (81%) | Lim. | **Lim. (11)**, Dep. (2), B.b (2) | D.sic | D.w (5), **D.sic (4)**, D.lx (4) |
| 6 | 13/14 (93%) | 9/13 (69%) | Dep. | **Dep. (6)**, Lim. (6) | D.w | D.lx (8), **D.w (6)** |
| 7 | 15/18 (83%) | 12/17 (71%) | Dep. | B.b (6), **Dep. (5)**, C.s (3) | D.lx | **D.lx (11)**, D.sic (3) |
| 8 | 8/12 (67%) | 7/12 (58%) | Lim. | Lim. (7), Dep. (2) | War. | D.sic (5), D.w (4), **War. (3)** |
| 9 | 13/15 (87%) | 12/14 (86%) | Lim. | **Lim. (9)**, B.b (3), Dep. (2) | D.lx | **D.lx (9)**, War. (2), D.w (2) |
| 10 | 9/12 (75%) | 8/12 (67%) | Lim. | Dep. (5), **Lim. (4)**, C.s (2) | D.sic | War. (5), **D.sic (4)**, D.w (2) |

*The relevant abbreviations for reason and purpose categories are as follows: Limitation (Lim.), Dependency (Dep.), Recursive call (R.c), Document (Doc.), Build break (B.b), Code smell (C.s), Document for later fix (D.lx), Document workaround (D.w), Warning for future developers (War.), Document suboptimal implementation choice (D.sic), Placeholder for later extension (Pla.). We only consider the labels that have more than one response as valid survey results.*

practically useful to have tools or systems to help practitioners manage SATD traceability automatically. A SATD management tool could make the developers aware of the debt being incurred and would make it easy to continually avoid the debt as part of their normal workflow. A possible mock-up was presented by Maipradit et al. [26, Fig. 7]. Especially for third-party libraries, an effective awareness mechanism is needed to allow upstream developers to continually fix library related issues. In RQ1, we find that *External library limitation* is the most frequently occurring reason subcategory (41%). Moreover, such limitation is also pointed by one survey response: "*This situation sucks, if some of these libraries just don't have a (maintained) osgi-ified bundle. This is one of those cases and there's not much you can do aside from bundle it yourself which can be error prone and a maintenance nightmare.*"

- *Focusing on top SATD locations and reasons would provide the most benefit to developers.* In RQ1, we provide the most frequently occurring locations and reasons for SATD in the build systems. We suggest that tool builders make an extra effort on these top locations and reasons.

- *Tool support for recommending solutions to SATD in build systems.* During the creation of pull requests and issue reports, we observe that the possible solutions that we provided for developers to mitigate 'ready-to-be-addressed' SATD are similar and straightforward (e.g., remove the extraneous comment or code). This observation suggests that an automated tool for addressing SATD could be useful. This will not just help developers to manage such SATD, but also will improve the quality of the final product.

## 8 RELATED WORK

In this section, we position our work with respect to the literature on build systems and technical debt related to this study.

### 8.1 Build System

Build system maintenance is a hidden cost, which takes a considerable amount of development effort. Kumfert et al. [22] argued that the need to keep the build system synchronized with the source code generates an implicit overhead on the development process, and in their survey, developers claimed that they spend up to 35.71% of their time on build system maintenance. McIntosh et al. [32] analyzed ten large, long-live projects by mining the version histories, and their study showed that build system maintenance is 27% overhead on source code development. Adams et al. [1] studied the evolution of the Linux KBUILD files and how these files co-evolve with the source code. McIntosh et al. [31] made similar observations in Java build systems.

Build breakage and how to repair it have been widely studied. Kerzazi et al. [20] interviewed 28 software engineers to study why build breakages are introduced in an industrial setting. Rausch et al. [39] performed an analysis of build failures, which studies the variety and frequency of build breakage in the CI environments of 14 open source Java projects. Islam and Zibran [19] studied the factors that may impact the build outcome, observing that the number of changed lines of code, files, and built commits in tasks are most significantly associated with build outcomes. Zolfagharinia et al. [53] studied the impact of operation system and runtime environment on build breakage in the CI environment of the Comprehensive Perl Archive Network (CPAN) ecosystem, suggesting interpretation of build results is not straightforward.

In addition, researchers have proposed automated approaches to repair build breakages. For example, Macho et al. [25] proposed BUILDMEDIC, an approach to automatically repair Maven builds that break due to dependency-related issues. Hassan and Wang [15] proposed HireBuild, an approach to automatically repair build scripts with fixing histories. Hassan [14] also outlined promising preliminary work towards automatic build repair in CI environment that involves both source code and build script.

There have also been other predictive approaches proposed to promote awareness and simplify interactions with build systems. Tufano et al. [46] envisioned a predictive model that would preemptively alert developers about the extent to which their software changes may impact future building activities. Hassan and Zhang [13] defined a model for predicting the certification results of a software build. Bisong et al. [6] proposed and analyzed models that can predict the build time of a job. Cao et al. [7] proposed BuildMétéo—a tool to forecast the duration of incremental build jobs by analyzing a timing-annotated Build Dependency Graph (BDG).

Although plenty of studies widely investigate the importance of build system maintenance and propose approaches to relieve the build issue, there is no study that focuses on SATD within the scope of build system maintenance. However, build systems often suffer from massive maintenance activities during the development process, and the part of these activities is produced by SATD, since SATD changes are more difficult to perform and SATD inevitably generate long-term maintenance problems from a short-term hack. Thus, in this study, we first characterize and mitigate SATD in the Maven build system and explore the feasibility of training automatic classifiers to identify SATD characteristics.

### 8.2 Technical Debt

Technical Debt is a design or implementation construct that is expedient in the short term, but sets up a technical context that can make a future change more costly or impossible [3]. Due to the importance of technical debt to the software development process and quality, there have been surveys and mapping studies about technical debt. Sierra *et al*. [45] surveyed research work on SATD, analyzing the characteristics of current approaches and techniques for SATD detection, comprehension, and repayment. Li *et al*. [23] performed a mapping study on technical debt and technical management. Vassallo *et al*. [47] showed that 88% of participants mentioned documenting their suboptimal implementation choices in the code that they produced.

Prior studies widely analyzed the factors or activities that affect technical debt. Besker *et al*. [4] observed that six organizational factors (experience of developers, software knowledge of startup founders, employee growth, uncertainty, lack of development process, and the autonomy of developers regarding TD decisions) were associated with the benefits and challenges of the intentional accumulation of technical debt in software. Besker *et al*. [5] also investigated activities on which wasted time is spent and whether different TD types impact the wasted time in different ways.

The detection of technical debt is also widely studied. Liu *et al*. [24] proposed the SATD detector to automatically detect SATD comments and highlight, list, and manage detected comments in an Integrated Development Environment (IDE). Farias *et al*. [10] carry out three empirical studies to curate the knowledge embedded in the SATD identification vocabulary, which can be used to automatically identify and classify TD items through code comment analysis. Yan *et al*. [50] also proposed an automated change-level TD determination model that can identify TD-introducing changes. Wattanakriengkrai *et al*. [49] combine N-gram IDF and auto-sklearn machine learning approaches to train classifiers to identify requirement and design debt. Maldonado *et al*. [29] used NLP maximum entropy classifiers [30] to automatically identify design and requirement SATD in source code comments. Moreover, Ren *et al*. [40] used Convolution Neural Network-based approaches with baseline text-mining approaches [18] to identify SATD in a cross-project prediction setting. Maipradit *et al*. [26], [27] identified "On-Hold" SATD for automated management.

Inspired by these past studies of SATD, in this paper, we conduct the first study on self-admitted technical debt in build systems. Similar to prior work, we first set out to characterize SATD in build systems in terms of locations, reasons, and purposes. We provide three coding guides for SATD in build systems, and automated SATD classifiers are provided in Section 4. Furthermore, in this work, we investigate the willingness of developers to remove the 'ready-to-be-addressed' SATD that refers to resolved issue reports.

## 9 THREATS TO VALIDITY

Below, we discuss the threats to the validity of our study:

*Construct Validity*. Construct validity is concerned with the degree to which our measurements capture what we aim to study. We use comment patterns to identify SATD comments in build files. Since SATD comment patterns are not enforced, we will miss SATD comments that do not conform to these comment patterns. To mitigate this risk, we expand upon a popular comment patterns list [38] with features recommended by Huang *et al*. [18]. Since we did not submit issues and pull requests for non-SATD changes to build systems, we do not have a baseline to which we can compare the prioritization of SATD issues and pull requests. Nevertheless, the proclivity of projects to accept SATD-related build changes (43% of PRs were accepted) and issue reports (56% were confirmed) is still a positive indication that these contributions are valuable to some degree. Future work is needed to gain a clearer impression of the relative importance.

*Content Validity*. Content validity is concerned with the degree to which a measure represents all facets of a given construct. In our study, we did not manually classify the whole dataset of SATD comments in build files, which will bring the risk of undiscovered SATD characteristic categories. Nonetheless, we strive for theoretical saturation [11] to achieve analytical generalization. Theoretical saturation is widely adopted in the SE domain [17], [41], [52]. To ensure that no new codes have been discovered, we performed four iterations and achieved saturation after coding 266 SATD comments.

*Internal Validity*. Internal validity is the approximate truth about inferences regarding cause-effect or causal relationships. We rely on manually coded data, which may be mis-coded due to the subjective nature of understanding the coding schema. To mitigate this threat, we apply three best practices for open coding: 1) we conduct four rounds of independent coding and calculate the Cohen's Kappa to ensure that our agreements at least are 'Substantial'; 2) we pursue saturation with concrete criteria, i.e., 50 consecutively coded comments for which no new categories are discovered; 3) we perform two passes that revisited miscoded SATD comments based on additional contextual information.

*External Validity*. External validity is concerned with our ability to generalize based on our results. We only conduct an empirical study of 291 Maven projects. As such, our results may not generalize to all Maven projects or other build technologies. On the other hand, our sample of projects is diverse, including projects of varying size and domain. Nonetheless, replication studies may help to improve the strength of generalizations that can be drawn.

## 10 CONCLUSION

Addressing self-admitted technical debt (SATD) is an important step in the development process. Recently, many studies have focused on SATD in source code, but little is known about SATD in the build systems. Thus, in this paper, we characterize and propose mitigation strategies for the SATD in build systems. To do that, we (i) manually classified 500 SATD comments according to their locations, reasons, and purposes; and (ii) trained SATD classifiers using the coded SATD comments; and (iii) investigated the willingness of developers to remove the 'ready-to-be-addressed' SATD that references resolved bug reports.

We observe that (i) SATD comments in Maven build systems most often occur in the plugin configuration location; the most frequently occurring reasons behind SATD is to document limitations in tools and libraries, as well as issues to be fixed later; and (ii) our auto-sklearn classifier achieves better performance than baseline classifiers, achieving an F1-score of 0.72–0.79; and (iii) the removal rates of 'ready-to-be-addressed' SATD in pull requests and issue reports reached 43% and 56%, respectively. We foresee many promising avenues for future work, such as improvements to the classifiers, expanding our coded corpus of SATD comments to other build systems, and automatic approaches to address SATD in build systems.

## ACKNOWLEDGMENTS

We would like to thank Rungroj Maipradit for providing technical assistance in training auto-sklearn classifier.

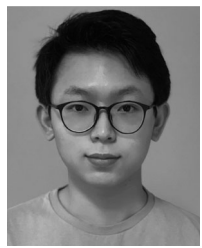

**Tao Xiao** is currently working toward the master's degree in the Department of Information Science, Nara Institute of Science and Technology, Ikoma, Japan. His main research interests include empirical software engineering, mining software repositories, natural language processing. For more information, please visit https://tao-xiao.github.io/.

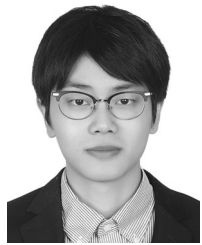

**Dong Wang** is currently working toward the doctor degree at the Nara Institute of Science and Technology, Ikoma, Japan. His research interests include code review and mining software repositories. For more information, please visit https://dong-w.github.io/.

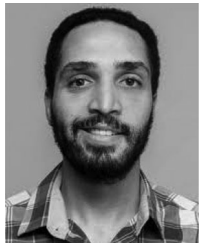

**Shane McIntosh** received the PhD degree from Queen's University, Kingston, Canada, for which he was awarded the Governor General's Academic Gold Medal. He is an associate professor with the University of Waterloo. Previously, he was an assistant professor with McGill University, where he held the Canada research chair in software release engineering. In his research, he uses empirical methods to study software build systems, release engineering, and software quality. For more information, please visit http://shanemcintosh.org/.

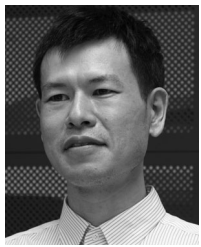

**Hideaki Hata** received the PhD degree in information science from Osaka University, Suita, Japan. He is an associate professor with Shinshu University. His research interests include software ecosystems, human capital in software engineering, and software economics. More about him and his work is available online at https://hideakihata.github.io/.

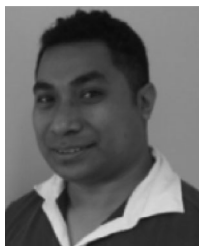

**Raula Gaikovina Kula** received the PhD degree from the Nara Institute of Science and Technology, Ikoma, Japan, in 2013. He is an assistant professor with the Nara Institute of Science and Technology. His interests include software libraries, software ecosystems, code reviews, and mining software repositories.

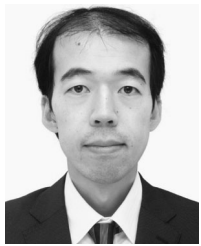

**Takashi Ishio** (Member, IEEE) received the PhD degree in information science and technology from Osaka University, Suita, Japan, in 2006. He was a JSPS research fellow from 2006–2007. He was an assistant professor with Osaka University from 2007–2017. He is now an associate professor of the Nara Institute of Science and Technology. His research interests include program analysis, program comprehension, and software reuse. He is a member of the ACM, IPSJ, and JSSST.

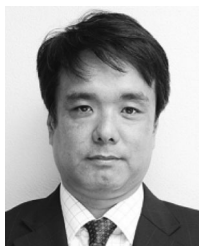

**Kenichi Matsumoto** (Senior Member, IEEE) received the BE, ME, and PhD degrees in engineering from Osaka University, Suita, Japan, in 1985, 1987, and 1990, respectively. He is currently a professor with the Graduate School of Information Science, Nara Institute Science and Technology, Japan. His research interests include software measurement and software process. He is a member of the IPSJ and SPM.

▷ **For more information on this or any other computing topic, please visit our Digital Library at** www.computer.org/csdl.